\documentclass[a4paper,11pt]{article}
\usepackage{aaskaiid}
\usepackage{graphicx}
\usepackage{soul}
\usepackage[ragged]{sidecap}    
\usepackage{sidecap}
\usepackage{ragged2e}
\usepackage[font=small]{caption}
\usepackage{orcidlink}

\title{The SKA View of Cool-core Clusters: \\
Evolution of Radio Mini-halos and AGN Feedback}
\ShortTitle{The SKA View of Cool-core Clusters}

\author[1,2]{Myriam Gitti\orcidlink{0000-0002-0843-3009}}
\ShortName{M. Gitti} 
\author[3]{Paolo Tozzi\orcidlink{0000-0003-3096-9966}}
\author[1,2]{Francesco Ubertosi\orcidlink{0000-0001-5338-4472}}
\author[1,2]{Annalisa Bonafede\orcidlink{0000-0002-5068-4581}}
\author[2]{Gianfranco Brunetti\orcidlink{0000-0003-4195-8613}}
\author[2]{Rossella Cassano\orcidlink{0000-0003-4046-0637}}
\author[4,5]{Stefano Ettori\orcidlink{0000-0003-4117-8617}}
\author[2]{Luigina Feretti\orcidlink{0000-0003-0312-6285}}
\author[6]{Marie-Lou Gendron-Marsolais\orcidlink{0000-0002-7326-5793}}
\author[7]{Simona Giacintucci\orcidlink{0000-0002-1634-9886}}
\author[8]{Julie Hlavacek-Larrondo\orcidlink{0000-0001-7271-7340}}
\author[9]{Alessandro Ignesti\orcidlink{0000-0003-1581-0092}}
\author[3,10] {Giulia Macario\orcidlink{0000-0003-3985-5105}}
\author[11]{Mamta Pandey-Pommier\orcidlink{0000-0001-5829-1099}}
\author[2]{Tiziana Venturi\orcidlink{0000-0002-8476-6307}}
\affiliation[1]{Dipartimento di Fisica e Astronomia, Universit{\`a} di Bologna, via Gobetti 93/2, 40129 Bologna, Italy}
\affiliation[2]{INAF -- Istituto di
Radioastronomia, via Gobetti 101, 40129 Bologna, Italy}
\affiliation[3]{INAF -- Osservatorio Astrofisico di Arcetri, Largo Enrico Fermi 5, 50125 Firenze, Italy}
\affiliation[4]{INAF -- Osservatorio di Astrofisica e Scienza dello Spazio, via Gobetti 93/3, 40129 Bologna, Italy}
\affiliation[5]{INFN -- Sezione di Bologna, viale Berti Pichat 6/2, 40127 Bologna, Italy}
\affiliation[6]{Dép. de physique, de génie physique et d’optique, Université Laval, Québec (QC), G1V 0A6, Canada}
\affiliation[7]{Naval Research Laboratory, 
4555 Overlook Avenue SW, Code 7213, 
Washington, DC 20375, USA}
\affiliation[8]{Département de physique, Physics department, Université de Montréal}
\affiliation[9]{INAF -- Astronomical Observatory of Padova, vicolo dell'Osservatorio 5, IT-35122 Padova, Italy}
\affiliation[10]{SKAO, SKA-Low Science Operations Centre, 26 Dick Perry Avenue, Kensington WA 6151, Australia}
\affiliation[11]{Pole Scientific, University Catholic of Lyon- University of Lyon, 10 place des Archives 69288, Lyon, France}
\emailAdd{myriam.gitti@unibo.it}

\abstract{ 
In about 70\% of relaxed, cool-core galaxy clusters, the brightest cluster galaxy (BCG) is radio loud, showing non-thermal radio jets and lobes ejected by the central active galactic nucleus (AGN).  In recent years such relativistic plasma has been shown to interact with the surrounding thermal intra-cluster medium (ICM) as revealed by striking images where radio lobe fill the cavities in the X-ray-emitting gas.  This `radio-mode feedback' phenomenon is widespread and crucial for understanding the physics of cluster cores and the properties of the central BCG. Mechanically-powerful AGN are expected to drive turbulence in the central ICM which may also contribute to the origin of non-thermal emission on cluster-scales.  Diffuse non-thermal emission has been observed in many cool-core clusters in the form of a radio mini-halo surrounding the radio-loud BCG on scales comparable to the cooling radius.  Large samples of mini-halos are essential to clarify their origin and their link with the thermal and dynamical properties of clusters, especially in view of future high-resolution X-ray studies with {\it NewAthena} X–IFU.  All-sky surveys with the SKA-Mid telescope at arcsecond resolution would have the potential to detect up to $\sim$3500 mini-halos at redshift $z<1$ (compared to the few tens currently known). Deep Tier surveys with the SKA-Mid at sub-arcsecond resolution would further enable a complete census of radio-loud BCGs down to 1.4 GHz powers of $10^{23}$ W/Hz up to $z \sim 2$ . This will provide a comprehensive view of AGN feedback and its role in shaping large scale structures.
}

\begin{document}
\maketitle

\section{Introduction}\label{intro.sec}

\begin{SCfigure*}[0.7][t]
  \includegraphics[width=0.55\linewidth]{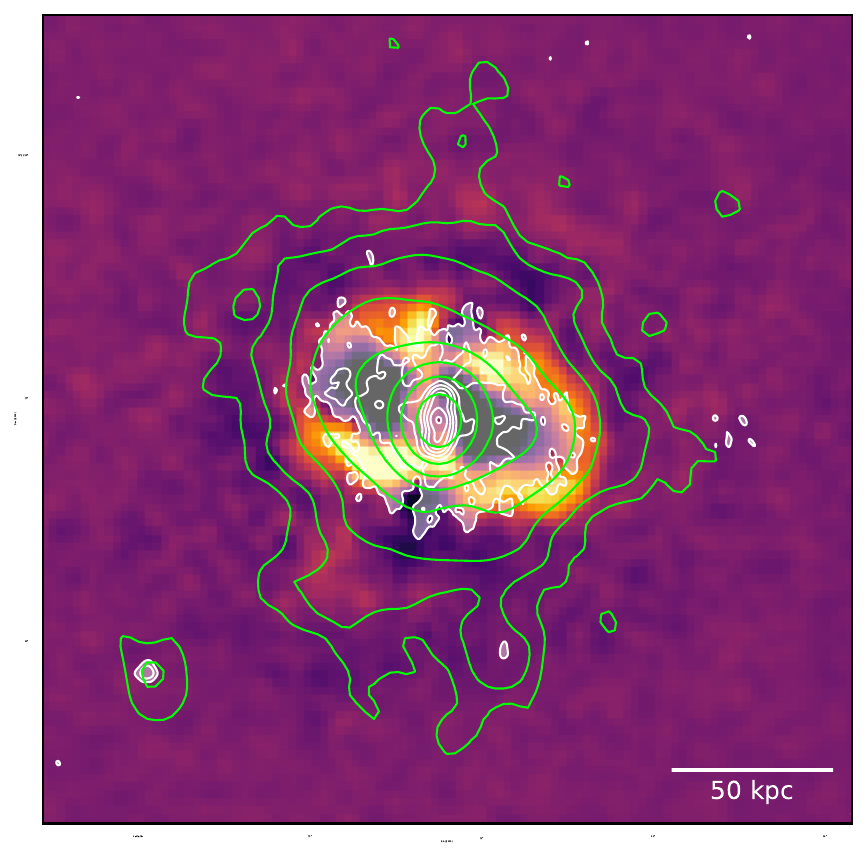}
    \caption{\justifying The X-ray {\it Chandra} image (residual after $\beta$-model subtraction) of the galaxy cluster RBS~797 at $z = 0.35$ \citep{Ubertosi_2021} is shown in color scale with superimposed VLA contours at 1.4 GHz \citep[green,][]{Gitti_2006} and at 3~GHz \citep[white,][]{Ubertosi_2024a}. 
    Contours levels start at 3$\times$rms (rms $\sim$13~$\mu$Jy/beam at 1.4 GHz and 5~$\mu$Jy/beam at 3 GHz; beam $\sim$ 3$''$ at 1.4 GHz and 0.9$''$ at 3 GHz), and increase by a factor of 2. The X-ray cavities, visible as depressions in the X-ray surface brightness, are filled with the BCG radio lobes (white contours), that have also driven cocoon shock fronts in the ICM \citep{Ubertosi_2023b}; on larger scale, a radio mini-halo is detected (green contours). }\label{RBS797feedback.fig}
\end{SCfigure*}

The majority of baryons in galaxy clusters reside in the hot, diffuse intra-cluster medium (ICM), mostly heated by gravitational processes to 1-10 keV and observable via thermal bremsstrahlung and line emission in X-rays. Yet, the expected catastrophic cooling and the consequent formation of adiabatic cooling flows in relaxed cluster cores \citep{Fabian-CF_1994} is not observed \citep[e.g.,][]{Peterson-Fabian_2006}, implying that non-gravitational heating processes act on $\sim$100 kpc scales around the brightest cluster galaxy (BCG). Understanding the interaction between gravitational physics, relativistic plasma injected by the active galactic nucleus (AGN), and the ICM is key to tracing galaxy growth, star formation history, and large-scale structure assembly.
In particular, AGN feedback has emerged as a fundamental component of galaxy evolution models, with its clearest manifestation in {\it cool-core} (CC) clusters -- systems with short central cooling times, high densities ($n_{\rm e} > 10^{-2}$ cm$^{-3}$), and relatively low temperatures ($kT <$ few keV). 
Although the presence of ICM at temperatures below $\sim$1 keV, as predicted by the classical adiabatic cooling-flow model, is generally not observed \citep[e.g.,][]{Peterson-Fabian_2006} except in a few rare cases \citep{Pinto_2018}, the strength of the cool core has nonetheless been shown to correlate with both the star formation rate and the nuclear accretion rate, suggesting that systems with the strongest cool cores host cooling processes \citep{McDonald_2018}.
Therefore, CC clusters  are ideal laboratories for studying AGN–ICM interactions.
\\
BCGs in CC clusters often host FR-I radio sources \citep{Fanaroff-Riley_1974}, detected in $\sim$70\% of cases \citep{Burns_1990, Dunn-Fabian_2006, Best_2007, Mittal_2009}. {\it Chandra} and {\it XMM-Newton} observations reveal that the X-ray-emitting gas is displaced by radio jets, which inflate lobes (radio {\it bubbles}) and create {\it cavities} visible as depressions in the X-ray surface brightness. Striking composite images, where radio lobes precisely fill X-ray cavities, establish radio galaxies as
a primary source of feedback able to mitigate and possibly solve the cooling flow problem \citep[e.g.,][]{McNamara-Nulsen_2007, Gitti_2012, Fabian_2012, Gaspari_2013, Donahue-Voit_2022}.
The mechanisms by which AGN jets heat the ICM likely involve buoyant bubbles, shocks, and turbulence \citep[e.g.,][]{Eckert_2021}.
AGN-inflated bubbles may dissipate enthalpy as they rise through the ICM \citep{McNamara-Nulsen_2007, McNamara-Nulsen_2012}, and are observed in roughly 50\% of mass-selected and 60-90\% of X-ray-selected CCs \citep{Olivares_2022, Fabian_2012}. Jets can also drive shocks \citep[e.g.,][]{Heinz_2005},
which may offer a more direct and isotropic energy transfer \citep{Nulsen-McNamara_2013, Ubertosi_2023b}. However, due to observational limits, to date fewer than twenty systems show clear detections of shocks \citep[e.g.,][]{Ubertosi_2023b}. 
A third mechanism is AGN-driven turbulence, which may both heat the ICM \citep[e.g.,][]{Zhuravleva_2014} and re-accelerate relativistic particles \citep{Bravi_2016}, possibly seeded by disrupted AGN bubbles.
\\
Diffuse non-thermal radio emission is indeed observed in a few tens of CC clusters as {\it radio mini-halos}, surrounding the BCG. These faint, steep-spectrum ($\alpha > 1$, where the flux density at the frequency
$\nu$ is $S_{\nu} \propto \nu^{-\alpha}$) sources extend over $\sim$100-500 kpc and have typical 1.4 GHz powers of $\sim10^{23}$-$10^{24}$ W/Hz \citep[e.g.,][]{Feretti_2012, vanWeeren_2019}. Unlike bubbles, mini-halos originate from the ICM itself, where thermal plasma and relativistic electrons mix. Figure~\ref{RBS797feedback.fig} shows a clear example of radio lobes (white contours), filling X-ray cavities, which are distinct from the mini-halo (green contours).
Mini-halos trace $\sim$GeV electrons in $\sim$$\mu$G magnetic fields, similar to giant halos.
However, while giant halos (which extend on $\gtrsim$ Mpc-scale) are merger-driven and found in non-CC clusters \citep[see e.g.,][]{Cassano01.2026.SKA}, mini-halos are detected in relaxed CC systems and are likely linked to AGN activity, though not directly powered by it. Electrons may be injected by the AGN and re-accelerated by turbulence in the core, possibly driven by the AGN itself \citep[e.g.,][]{Richard-Laferriere_2020} or gas sloshing \citep[e.g.,][]{Mazzotta-Giacintucci_2008}. This suggests that mechanisms less extreme than mergers, as AGN feedback and sloshing are, can sustain non-thermal emission (note however that while radio emission is always associated with the BCG in CC clusters, AGN feedback and sloshing are not ubiquitous, e.g., \cite{Giacintucci_2019} and references therein), though the exact process remains uncertain (see Sect.~\ref{origin.sec}).
\\
Cool cores are already established at $z\sim1$ \citep{Santos_2012}, and are consistently associated with both radio-loud AGN and mini-halos  \citep{Sun_2009b, Feretti_2012}.  However, current systematic studies of AGN feedback and mini-halos are mostly limited to $z < 0.7$ \citep{Hlavacek_2012}\footnote{One mini-halo candidate was recently identified at $z \sim 1.7$ \citep{Hlavacek_2025}.}, despite ICM in cluster halos is being detected up to $z \gtrsim 1.6$ \citep[e.g.,][]{paolo_2015, Mantz_2020, Hlavacek-Larrondo_2020, Andreon_2023, Lepore_2024}. A mild negative evolution 
of mini-halos with redshift may be associated to the decrease of CC clusters in favor of a population of mergers or dynamically unrelaxed clusters \citep{Santos_2010, McDonald_2013}. 
However, the current gap is mainly due to observational challenges: mini-halos must be disentangled from the bright central AGN emission, which can reach few Jy/arcsec$^{2}$ at 1.4~GHz (as e.g., in Perseus), while the typical mini-halo brightness is only a few $\mu$Jy/arcsec$^{2}$. This requires high sensitivity, dynamic range, and sub-arcsec to milliarcsec angular resolution — capabilities uniquely offered by the Square Kilometre Array Observatory\footnote{https://www.skao.int/en/science-users/118/ska-telescope-specifications} \citep[SKAO,][]{SKAPerformance_2019}.
\\
\indent
Studying non-thermal activity and radio-mode feedback is essential for understanding ICM physics and cosmic structure evolution. The SKAO will play a unique and game-changing role in tracing non-thermal emission associated to mini-halos and AGN feedback in cluster cores up to the highest redshifts.
In this chapter, we focus on insights into radio mini-halos (Sect.~\ref{MH.sec}) and AGN feedback (Sect.~\ref{feedback.sec}) from SKA-Mid statistical studies, updating the work presented in \cite{Gitti_2015} and highlighting the strong synergy with future X-ray missions. 
Potential contributions from SKA-Low surveys are expected to be similar to those discussed in previous mini-halo studies \citep{Gitti_2018}, while being less optimal for AGN feedback investigations due to the limited angular resolution, and are therefore only briefly discussed here.
Our conclusions are summarized in Sect.~\ref{conclusions.sec}.
Throughout the paper, we adopt a $\mathrm{\Lambda CDM}$ cosmology with $\mathrm{H_{0}=70}$
km $\mathrm{s^{-1}Mpc^{-1}}$, $\Omega_{\rm M} = 1 - \Omega_{\Lambda} =
0.3$.

\section{Radio mini-halos}
\label{MH.sec}

\subsection{Current understanding of the mini-halo origin and open questions }
\label{origin.sec}

Galaxy clusters hosting radio mini-halos invariably contain a central radio-loud AGN, often showing radio lobes and bubbles that inject relativistic electrons into the cluster core. While AGN may provide the primary source of relativistic particles, they cannot alone account for the diffuse emission without additional dynamical input.
As with giant radio halos \citep[see also][]{Cassano01.2026.SKA}, the origin of mini-halos is challenged by the short radiative lifetime of relativistic electrons compared to the size of the emission region. Two main models address this "slow-diffusion problem": {\it \textup{in situ}} re-acceleration by turbulence in the cool core \citep[][]{Gitti_2002, Gitti_2004, Mazzotta-Giacintucci_2008, ZuHone_2013}, and hadronic models where secondary electrons are produced via $p$-$p$ collisions \citep[][]{Pfrommer-Ensslin_2004, Jacob-Pfrommer_2017b, Ignesti_2020}; see \citet{Brunetti-Jones_2014} for a review.
\\
\indent
A key question in re-acceleration models is the origin of turbulence and its link to the central AGN. Turbulence may arise from AGN jets and lobes interacting with the ICM \citep[e.g.,][]{Heinz_2006, Zhuravleva_2016}.
Turbulent heating has also been proposed as a mechanism for maintaining thermal balance in the ICM \citep{Dennis-Chandran_2005, Gaspari_2012, Zhuravleva_2014}. If turbulence is partly dissipated into heat and partly channeled into particle re-acceleration, a shared origin for both gas heating and mini-halos is possible \citep{Bravi_2016}. 
A decade ago, the first direct measurement of turbulence was obtained in the Perseus cluster, and it has been shown to be consistent with re-acceleration requirements \citep{Hitomi_2016, Hitomi_2017}.
With the advent of the X-Ray Imaging and Spectroscopy Mission (\emph{XRISM}; \citealp{XRISM_quick}), launched in 2023, turbulence has been measured with the spectrometer {\it Resolve} in a handful of nearby clusters, providing an observational basis for mini-halo models
\citep{XRISM-Perseus_2025, XRISM_A2029_2025, XRISM_2319_2025}.  
However, current measurements provide contradictory results on the consistency with the expected level of turbulence from hydrodynamic simulations \citep[e.g.,][]{Vazza_2025}.
This scenario implies a close but complex link between thermal and non-thermal components, which requires to be further tested in the case of AGN-induced turbulence by examining AGN feedback in mini-halo clusters (see Sect.~\ref{feedback.sec}). With future generation X-ray facilities, this 
scenario can be investigated in greater detail with spatially resolved analysis (see Sect.~\ref{athena.sec}).
Hadronic models offer a viable alternative, where cosmic-ray instabilities heat the ICM \citep{Fujita-Ohira_2012, Jacob-Pfrommer_2017b}. Unlike giant halos, the expected $\gamma$-ray emission from mini-halos remains consistent with current observational limits \citep[e.g.,][]{Ahnen_2016, Jacob-Pfrommer_2017b, Ignesti_2020}, leaving this alternative model for the origin of mini-halos substantially unconstrained.
\\
\indent
Another open issue is the link between mini-halos and cluster gas dynamics. Mild dynamical activity in some systems suggests that, beyond AGN feedback, minor mergers and sloshing may also inject turbulence \citep[e.g.,][]{Gitti_2007b, Cassano_2008b, Mazzotta-Giacintucci_2008, Riseley_2022}. Simulations support this view \citep[e.g.,][]{ZuHone_2013}, as do spiral-shaped cold fronts in CC clusters \citep[e.g.,][]{Markevitch-Vikhlinin_2007}, likely caused by subhalo-induced sloshing \citep{Fujita_2004, Ascasibar_2006, ZuHone_2016}.
A hybrid scenario is also plausible, where AGN feedback drives most of the turbulence while sloshing mostly shapes the mini-halo morphology \citep{Richard-Laferriere_2020}. Furthermore, 
LOFAR has recently revealed a correlation between the radial radio spectral index and AGN-driven shock fronts in one system, showing a flattening trend from the center to the edge that is unique among mini-halos and reminiscent of shock-reaccelerated particles \citep{Bonafede_2023}.
However, the authors show that shock re-acceleration alone cannot reproduce the observed radial brightness profile of the mini-halo. One possibility is that AGN-driven shocks propagate through a pre-existing mini-halo, re-accelerating and compressing the radio plasma and thereby producing the observed radial spectral flattening. Alternatively, in the presence of a high magnetic field (with central value $>15 \, \mu$G and a radial declining profile), turbulent re-acceleration may also account for the observed radial spectral flattening, independently of shocks. In this case, similar behaviour would be expected in all mini-halos with comparable magnetic field strengths. Deeper observations and systematic studies of other mini-halos will be required to confirm or rule out this scenario.
\\
\indent
Some CC clusters also host large-scale radio emission in the form of giant radio halos or hybrid morphologies \citep[e.g.,][]{Sommer_2017, Savini_2019, Biava_2021b, Biava_2024, Riseley_2022, Lusetti_2024}. 
The Perseus cluster, for example, shows filamentary substructures and a distinct large-scale radio halo \citep{vanWeeren_2024}. 
These sources suggest that minor mergers, while not energetic enough to disrupt the cool core, may still trigger particle acceleration in the ICM on scales of hundreds of kiloparsecs.
A joint X-ray and radio analysis of a sample of CC clusters observed at a low frequency further indicates that cluster-scale diffuse radio emission is correlated to the presence of cold fronts  \citep{Biava_2024}. 
The co-existence of cluster-scale diffuse radio emission and cold fronts in these CC systems implies the presence of sloshing motions capable of generating sufficient turbulence to re-accelerate particles on large scales while preserving the central cool core.
The large-scale halo typically shows properties that are distinct from that of the inner mini-halo, indicating that these radio phenomena may coexist in the same cluster and suggesting a possible evolutionary link \citep[][]{Zandanel_2014, Brunetti-Jones_2014, vanWeeren-2026}.
\\
\indent
Our understanding of mini-halos thus remains incomplete. Distinguishing between the leptonic and hadronic origin of mini-halos, or understanding their connection with
AGN-induced turbulence and with the ICM dynamics, offers a unique probe of ICM microphysics, including magnetic field amplification and cosmic-ray transport. 
A breakthrough is expected from future SKA spectral and polarimetric observations, especially when combined with hard X-ray and $\gamma$-ray constraints. 

\subsection{Detecting radio mini-halos: radio follow-up of X-ray cluster samples }
\label{followup.sec}

Large-sample studies will be crucial to address key questions:
{\it{(i)}} Are mini-halos ubiquitous in CC clusters, and how does their occurrence evolve with redshift?  
{\it{(ii)}} What role does the central AGN play, and how often is AGN feedback observed in mini-halo clusters?  
{\it{(iii)}} 
What is the origin and role of turbulence in powering mini-halos?
We will show that surveys  with the SKA telescopes have the capability to address these fundamental questions.

The detection of radio mini-halos is complicated by the need to
separate their faint, low surface brightness emission from the bright
emission of the radio-loud BCG embedded in it (see Sect. \ref{intro.sec}).  This requires
observations performed with a technical setup ensuring high
sensitivity to large-scale, low surface brightness emission, provided by short baselines, but at
the same time allowing an accurate subtraction of discrete sources,
which are best identified by long baselines.
This may be the reason for the relatively low number of mini-halos known 
-- currently, about thirty mini-halos (or candidates)
have been detected \citep[e.g.,][]{Richard-Laferriere_2020}.
In fact, although a correspondence between radio mini-halos and cool cores is well recognized, it is still not clear
whether these radio sources are intrinsically common or rare in CC clusters.
Studies of mass-selected sample of massive clusters ($M_{\rm 500} > 6 \times 10^{14}$ M$_{\odot}$ derived from Planck observations) found an occurrence of mini-halos as high as 80\% \citep{Giacintucci_2017}. However, larger statistical samples, that include lower-mass systems, are crucial to firmly establish the occurrence of mini-halos among the general CC population.
\begin{figure*}
\centerline{
\includegraphics[scale=0.35, clip, trim={0 0 0 5cm}]{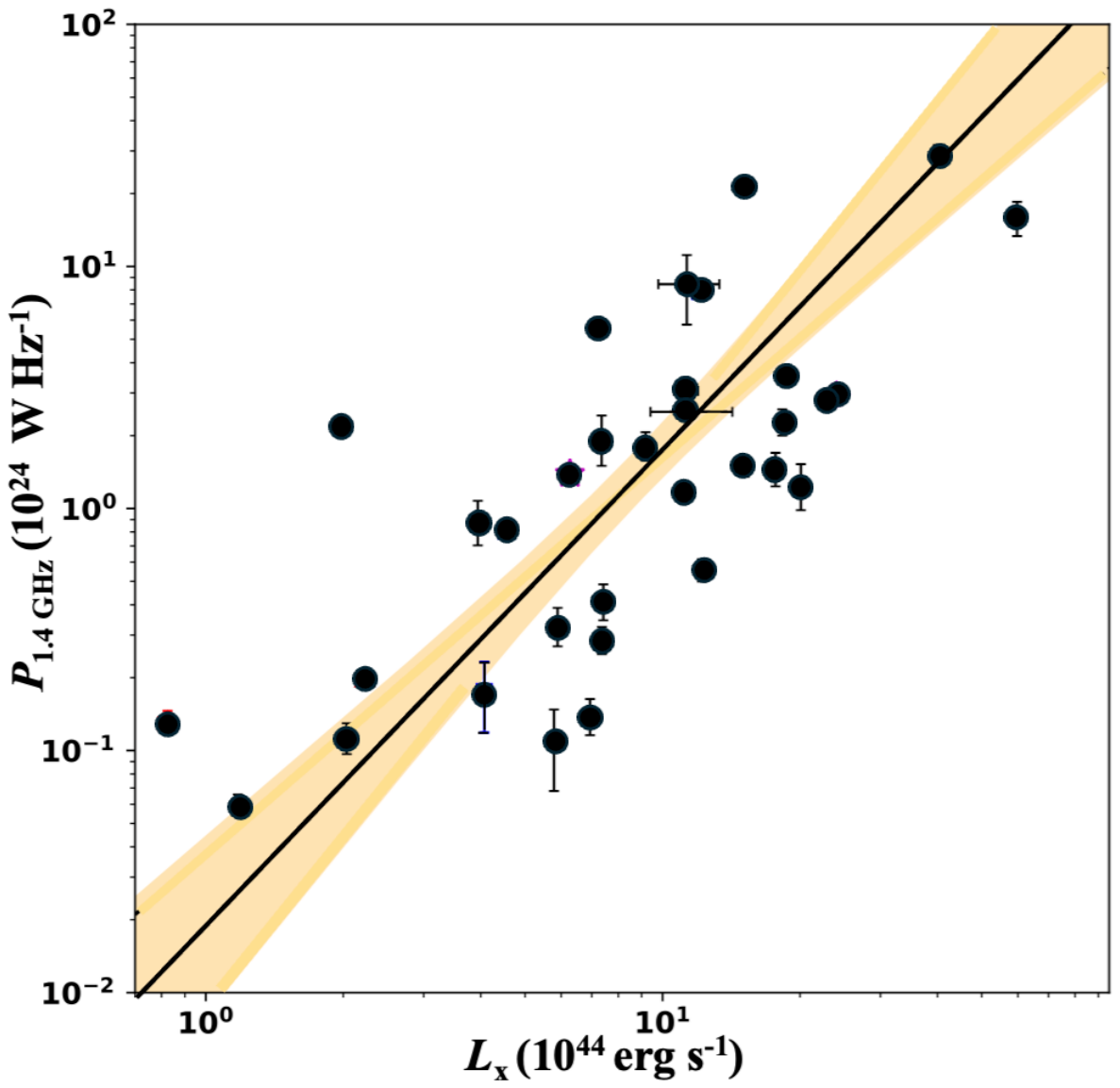}
\includegraphics[scale=0.35, clip, trim={0 0 0 2cm}]{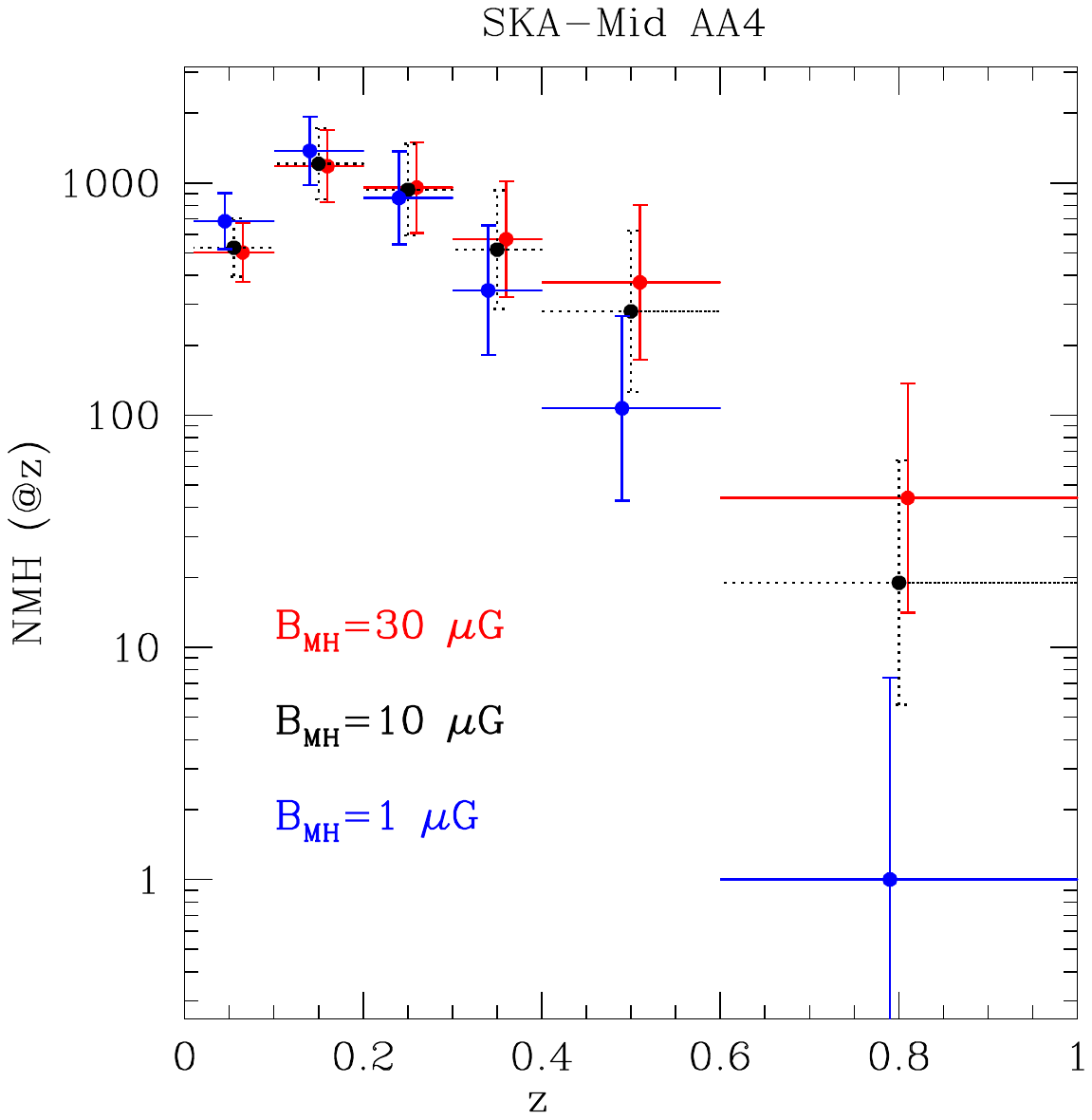}
\vspace{-2.cm}
}
\caption{\label{corr-NMH.fig} {\it Left panel:} Mini-halo radio power at 1.4~GHz versus the cluster X-ray luminosity inside a radius of 600 kpc for the mini-halo sample of \citet{Richard-Laferriere_2020}. The best-fitting line using the BCES-orthogonal method \citep{Akritas_1996}, which is calculated including also candidate and uncertain mini-halos, is displayed (black solid line) along with the 95\% confidence region (orange shaded area). The best-fit values are $a = 1.97 \pm 0.31$ and $ b = - 1.72 \pm 0.31 $ (see Eq. \ref{Pradio-Lx.eq}). 
{\it Right panel:} number of radio mini-halo expected in  
the all-sky ($3 \pi$ sr) survey by AA4 SKA-Mid at  Band 2 (see Table \ref{survey.tab})
in different redshift bins, estimated by assuming three reference values of magnetic field strength $B_{\rm MH}=1\,\mu$G, 10 $\mu$G, 30 $\mu$G (blue, black and red, respectively). The vertical error bars indicate the 1$\sigma$ uncertainty driven by the $P_{\rm 1.4}$-$L_{\rm X}$ correlation and the horizontal bars indicate the size of the redshift bin. The points corresponding to the different $B_{\rm MH}$-field values in the same redshift bin are slightly shifted on the $x$-axis to improve clarity. }
\end{figure*}
Good cluster statistics in terms of X-ray properties are already currently
available from {\it Chandra} and {\it XMM-Newton} and can be exploited
to forecast future detection of radio mini-halos, provided that an
intrinsic relation between the thermal and non-thermal cluster
properties exists.  For this purpose, \citet{Gitti_2015} and \citet{Gitti_2018}
investigated the link between
global X-ray and radio observables that can be easily obtained from
(present and future) cluster surveys, such as the luminosities.
Since all known mini-halos are hosted by clusters with low central
entropy \citep{Hudson_2010}, the basic assumption is that every
strong CC cluster hosts a radio mini-halo that follows the
observed correlation between the 1.4~GHz radio power of the
mini-halos, $P_{1.4}$, and the
global X-ray luminosity of the host clusters,
$L_{\rm X} $
\citep[e.g.,][]{Cassano_2008b, Kale_2013, Gitti_2018, Giacintucci_2019}.
This $P_{1.4}$-$L_X$ correlation can we written in the form:
\begin{equation}
\log P_{1.4} = a \, \log L_{\rm X} + b
\label{Pradio-Lx.eq},
\end{equation}
where $P_{1.4}$ is in units of $10^{24}$ W/Hz and 
$L_{\rm X} $ is in units of $10^{44}$ erg/s.
\\
\indent
In particular, \citet{Gitti_2018} estimated the maximum number of radio
mini-halo candidates that can potentially be discovered by the SKA-Mid and SKA-Low surveys as a function of redshift $z$.
This was achieved by integrating the radio luminosity function (RLF) 
of mini-halos over radio luminosity and redshift, where the RLF of
mini-halos is obtained from the X-ray luminosity function (XLF) 
of clusters\footnote{We considered the evolving XLF derived from the
high-redshift X-ray--selected 160 Square Degree {\it ROSAT} Cluster
Survey (160SD) by \cite{Mullis_2004} in a $\Lambda$-dominated
universe.} by considering the observed $P_{1.4}$-$L_X$ correlation and the
fraction of clusters with strong cool cores \citep[$\sim
0.4$,][]{Hudson_2010}.  
This correlation indicates a connection between the energy reservoir 
in CC clusters and that associated with the non-thermal components 
powering radio mini-halos \citep[see also][]{Bravi_2016}.
The non-thermal power, $P_{\rm NT}$, that can be sustained in the region of 
radio mini-halos over timescales longer than the radiative lifetime of relativistic 
electrons depends on the magnetic field strength as
$
P_{\rm NT} = P_{\rm radio} + P_{\rm IC} = P_{\rm radio} [1+ (
  B_{\rm CMB}/B_{\rm MH})^2 ],
$
where $P_{\rm radio}$ is the synchrotron power, $P_{\rm
    IC}$ is the inverse Compton (IC) power, $B_{\rm CMB} = 3.25 \,
(1+z)^2 \, \mu$G is the magnetic field equivalent to the inverse
Compton losses with CMB photons, and $B_{\rm MH}$ is the average magnetic field
intensity in the mini-halo region.
The observed synchrotron radiation is thus a function
of the intra-cluster magnetic field and redshift:~
$
P_{\rm radio}(B_{\rm MH},z) \propto B_{\rm MH}^2 / [B_{\rm MH}^2 + B^2_{\rm CMB}(z)] \, ,
$
where we made the dependencies explicit.
Conversely, by normalizing the radio luminosity to the observed X-ray luminosity 
through the measured $P_{1.4}$-$L_X$ correlation,  
the mini-halo RLF can be linked to the cluster XLF with a dependence on the parameters 
$a,b$ (see Eq. \ref{Pradio-Lx.eq}) and magnetic field $B_{\rm MH}$, only.
More details about this approach can be found in \citet[][and references therein]{Gitti_2018}. 
\\
In this chapter we update the previous work at mid frequencies by considering the $P_{1.4}$-$L_X$ 
correlation derived from the mini-halo compilation of \citet{Richard-Laferriere_2020},
with $a = 1.97 \pm 0.31$ and $ b = - 1.72 \pm 0.31 $ (see Fig. \ref{corr-NMH.fig}, left panel). 
We consider the radio observational performances 
of the full Design Baseline (Array Assembly 4, AA4) of the SKA-Mid telescope at Band 2 with rms of $\sim 4$ $\mu$Jy/beam at $\sim$8$''$ resolution (Briggs 0 image weighting and $\sim 4''$ tapering). According to the SKAO online sensitivity calculator\footnote{\url{https://sensitivity-calculator.skao.int}}, this sensitivity can be achieved with 10 minutes of on-source observing time,
thus a survey as wide as an all-sky survey ($3 \pi$ sr) could be completed in about 7 months. The survey details are summarized in Table \ref{survey.tab} (raw labeled "Mini-halos"). We note that this survey would also provide higher-resolution images ($< 1''$), which can be used to estimate and subtract the contribution of the BCG (raw labeled "(BCG subtract.)" in Table \ref{survey.tab}) from the extended radio emission.
Our predictions are shown in the right panel
of Fig. \ref{corr-NMH.fig}, where we plot the number of mini-halos expected
to be observable in different redshift bins, 
assuming three reference values of $B_{\rm MH} =1\, \mu$G (blue), 10 $\mu$G (black), and 30 $\mu$G (red).
Despite the large error bars, we note that models with low and high magnetic field values yield different expectations at $z > 0.6$. Specifically, the predicted number of mini-halos is only a few 
for $B_{\mathrm{MH}} = 1\,\mu\mathrm{G}$, and $\geq 20$ for $B_{\mathrm{MH}} = 30\,\mu\mathrm{G}$.
This is primarily due to the fact that, in the presence of weak magnetic fields, 
an increasing fraction of the non-thermal luminosity is channeled into IC emission 
at higher redshifts. As a result, mini-halos become progressively fainter and more challenging to detect.
Based on our estimates, upcoming SKA-Mid surveys will thus offer a mean to discriminate between different values of magnetic field strength in mini-halos. \\
We note that the predictions presented above for the AA4 SKA-Mid telescope, computed using the updated $P_{1.4}$–$L_X$ correlation \citep{Richard-Laferriere_2020}, are consistent with the results of \citet{Gitti_2018}, where the potential contributions from SKA-Low were also discussed. Accounting for the additional uncertainties in the spectral index when extrapolating to low frequencies, we therefore expect the predictions shown in Fig. 5 (bottom right panel) of \citet{Gitti_2018} to be valid also for the AA4 SKA-Low telescope. In particular, an all-sky survey with SKA-Low may be able to detect up to $10^4$ new mini-halos out to redshift $z \sim 1$, where a non-detection of mini-halos above $z\sim0.6$ would indicate a low magnetic field value $B_{\rm MH} < 1$ $\mu$G. We estimated that the observational performances adopted in that work (rms $\sim 20$ $\mu$Jy/beam at $\sim$ 8$''$ resolution, with Briggs 0 image weighting) can be achieved with 10 minutes of on-source observing time with the AA4 SKA-Low telescope, where higher resolution images ($\sim 3''$, with uniform weighting) can be used to account for the central BCG contribution. Considering the expected SKA-Low field of view (FOV $\sim 7$ deg$^2$ at the central frequency of 200 MHz), this survey could thus be completed in about 1 month. 

\subsection{Synergy with NewAthena: investigating the role of turbulent re-acceleration }
\label{athena.sec}
The estimates presented in Sect. \ref{followup.sec} assume that all CC clusters host a mini-halo, regardless of their mass or dynamical state. This may be reasonable within hadronic models, where the radio emissivity depends on the density of ICM thermal protons, but in re-acceleration or leptonic scenarios, sustaining radio-emitting electrons depends also on the turbulence level and the fraction of turbulent energy dissipated into particle re-acceleration.
As discussed in Sect. \ref{origin.sec}, turbulence may be driven by the interplay between outflowing relativistic plasma from AGN jets and lobes, responsible for AGN feedback, and mild dynamical activity inducing motions of the ICM.
If turbulence is triggered by core perturbations from large-scale dynamics \citep[e.g.,][]{ZuHone_2013}, it is expected to weaken in less massive and less disturbed systems. This may reduce the occurrence of mini-halos in CC clusters, leading to a lower number density than predicted by simplified models. On the other hand, turbulence may be higher at high redshift where clusters are still forming \citep[][]{Hlavacek_2025}.
Although this adds uncertainty to theoretical expectations, it underscores the importance of comparing them with future survey data. Such comparisons will help to constrain the role of turbulence and re-acceleration in mini-halos, and may clarify their physical origin.
According to numerical simulations \citep{ZuHone_2013}, turbulence with velocity broadening $\sigma_{\rm turb} \sim$ 50-200 km/s on spatial scales below $\sim$50-100 kpc in the CC region is potentially sufficient to re-accelerate seed relativistic electrons producing radio mini-halos. 
Such turbulence levels are in the reach of the X-IFU instrument on board {\it NewAthena}\footnote{\url{http://www.the-athena-x-ray-observatory.eu/}, expected launch date in late 2030s.}, which is expected to resolve gas velocities of $\sim$10 km/s in the cluster cores on $\sim$9 arcsec scales\footnote{9 arcsec corresponds to $\sim$70 kpc at z$\sim$1.} \citep{newathena_2025}.
\begin{figure*}
\centerline{
\includegraphics[scale=0.35, clip, trim={0 0 0 5cm}]{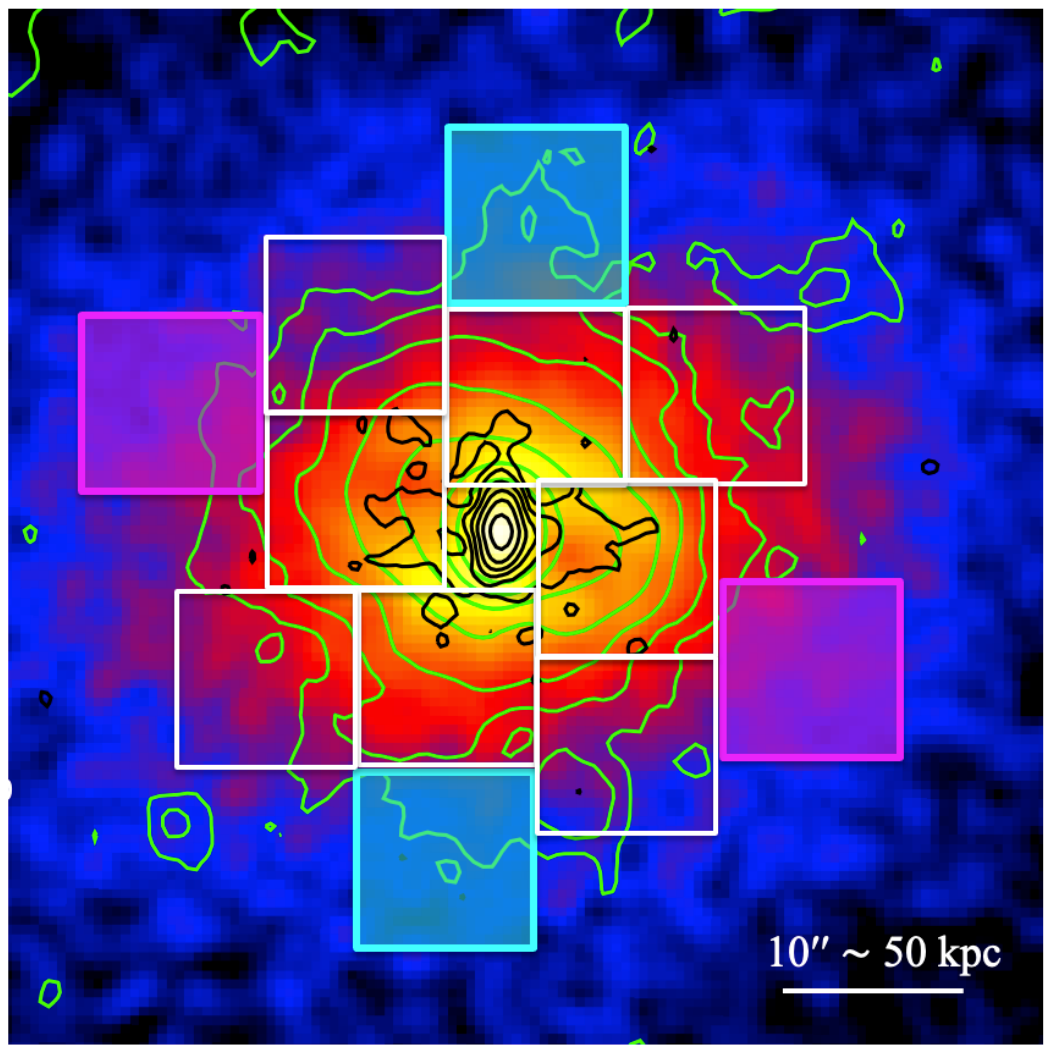}
\includegraphics[scale=0.35, clip, trim={0 0 0 5cm}]{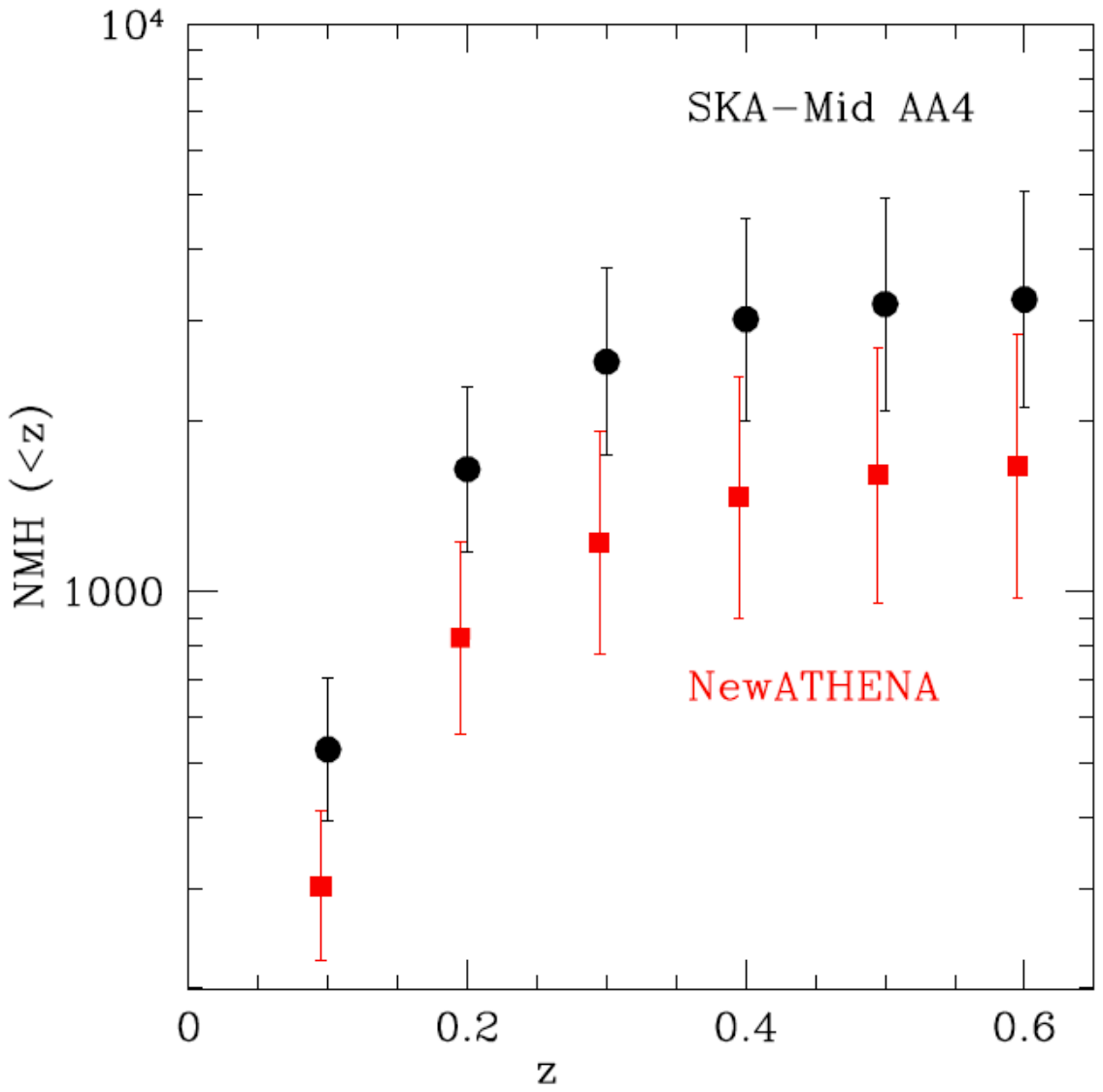}
\vspace{-2.cm}
}
\caption{\label{ska-athena.fig} {\it Left panel:}  {\it Chandra} image of the cluster RBS~797 (z=0.35, $1'' \sim$ 5 kpc) with the contours of the mini-halo radio emission \cite[in green, starting at 0.035 mJy/beam and increasing by a factor of 2, beam $\sim 3''$,][]{Doria_2012} and radio-loud BCG \citep[in black, starting at 0.03 mJy/beam and increasing by a factor of 2, beam $\sim$1.3$''$,][]{Gitti_2013a}. 
Shown in white is an example of the mesh for the point-to-point analysis, where the side of the box regions is $10''$, similar to the X-IFU pixel size. The shaded boxes are the ones we considered for a preliminary feasibility, with colors indicating the input value of the turbulent velocity broadening (cyan=70 km/s, magenta=120 km/s). By using the {\it Xspec} command {\ttfamily {fakeit}}, we simulated X-IFU spectra with an exposure of 2~Ms. 
{\it Right panel:} In black is shown the integral number of
  radio mini-halo candidates
detectable in the all-sky ($3 \pi$ sr) survey by AA4 SKA-Mid at Band 2 (see Table \ref{survey.tab})
up to a certain redshift, estimated by assuming a
  reference value of $B_{\rm MH}=10\, \mu$G (black). In red is shown the relative number of clusters that can be followed-up by {\it NewAthena} X-IFU with a point-to-point analysis measuring the turbulent velocity broadening in box regions (similar to the ones shown in the left panel), as described in the text.
  The error bars indicate the 1$\sigma$ uncertainty driven by the
  $P_{\rm 1.4}$-$L_{\rm X}$ correlation (Eq. \ref{Pradio-Lx.eq}). }
\end{figure*}
\\
\indent
Motivated by these considerations, we investigated the science capabilities possible with a synergy of SKA-Mid surveys and {\it NewAthena} to address the issue on the origin and role of turbulence in powering mini-halos.  
Point-to-point studies of the correlation between the radio and X-ray surface brightness have been proven to be important in investigating the origin of mini-halos \citep[e.g.,][]{Ignesti_2020}. The step forward we propose here is to adopt a similar approach to investigate the point-to-point correlation between the non-thermal properties of mini-halos and the ICM turbulent motions. This is essential to investigate the turbulent re-acceleration scenario for the origin mini-halos, and to evaluate the role of AGN feedback in powering the mini-halo emission.  
\\
\indent
We performed a preliminary feasibility study with X-IFU in a medium-redshift cluster, which shows clear bipolar radio lobes filling prominent X-ray cavities embedded in a mini-halo \citep[RBS~797 at z=0.35,][see Figs. \ref{RBS797feedback.fig} and \ref{ska-athena.fig}]{Gitti_2006,Gitti_2013a,Doria_2012,Ubertosi_2021}.
In particular, we simulated 2 Ms spectra in four $10'' \times 10''$ box regions along the edge of the halo emission (at radial distances $\sim 20''$), located in opposite directions so as to probe anisotropic Gaussian $\sigma_{\rm turb}$ for the turbulent velocity broadening (Fig. \ref{ska-athena.fig}, left panel).
By starting from the observed {\it Chandra} spectra, we derived the initial parameters of the spectral fit (temperature, abundance, normalization) and used these information to produce emission models which assume a turbulent velocity broadening with an anisotropic radial gradient, arbitrarily chosen to be higher along the AGN lobe direction as can be expected if the turbulence is induced by the AGN activity. Input values \citep[consistent with observational measurements, e.g.,][]{XRISM-Perseus_2025} 
range from 120 km/s along the AGN lobe – X-ray cavity direction to 70 km/s along the edge of the mini-halo in the opposite direction (magenta boxes and cyan boxes, respectively, in Fig. \ref{ska-athena.fig}, left panel).
By performing several {\ttfamily{Xspec}} simulations, we found that 
velocity broadening differences of approximately 70~km/s between various radial directions can be detected at $\approx 4\sigma$ on scales of $\sim$10$''$.
With this kind of study, we will thus be able to derive not only a global level of turbulence in CC clusters, providing a basis for comparison with more massive and dynamically active systems, but also the local correlation between turbulent velocity broadening and the non-thermal properties of the mini-halo. This, in turn, will help pinpoint the role of turbulence and re-acceleration mechanisms in mini-halos, and may offer a pathway to distinguish between different physical origins of these radio sources.
\\
We further estimated the integral number of clusters up to a certain redshift for which {\it NewAthena} can perform this kind of point-to-point analysis on SKA-detected mini-halo samples.
This is shown in the right panel of Fig.~\ref{ska-athena.fig} in comparison with the number of mini-halo candidates observable by SKA-Mid, assuming a reference magnetic field value of $B_{\rm MH}=10 \,\mu$G.
In particular, all-sky surveys conducted with the AA4 SKA-Mid configuration would 
be able to detect up to $\sim$3500 new mini-halos out to redshift $z \sim 0.6$ (and up to $10^4$ with the SKA-Low, see Sect. \ref{followup.sec}), 
thus producing a breakthrough in the study of these sources.  
We estimated that {\it NewAthena} can, in principle, perform a detailed follow-up of more than 50\% of SKA-detected mini-halos at any redshift, although each cluster would require a pointed exposure as we simulated. The best strategy will thus be to define well-selected, pilot sub-samples of mini-halo clusters \citep[as done e.g., in][]{Ignesti_2020} to undertake this kind of investigation.

\section{BCG radio properties and AGN feedback as a function of cosmic epoch}  
\label{feedback.sec}

\subsection{Tracing radio-mode feedback in cluster cores}
The radio-mode feedback is expected to be efficient in any cool core,
in order to prevent the formation of massive cooling flows 
and extreme star formation rates in the BCG \citep{McNamara_2006, Rafferty_2008}.
The presence of a radio galaxy in the BCG of a cluster is predicted to
be generally associated to radio bubbles (X-ray cavities) and shocks in the ICM (see Section~\ref{intro.sec}). As cavities are easier to detect than shock fronts, and hundreds of groups and clusters with cavities are known \citep[e.g.,][]{Shin_2016,Olivares_2022}, they are routinely used as tracers of AGN heating. The enthalpy of the cavities ($4 p V$ for relativistic plasma, where $p$ is the pressure of the surrounding ICM and $V$ is the cavity volume) is a proxy of the outburst energetics \citep{Birzan_2004, Birzan_2008}, which, in the few known cases of clusters with both cavities and shocks, has been shown to be comparable with the energetics of shock fronts \citep[e.g.,][]{Ubertosi_2023b} and scales with the X-ray radiative losses of the ICM \citep[e.g.,][]{Eckert_2021}. Radio BCGs are thus critical to maintain the large-scale cooling and heating balance and efficiently quench star formation in cluster cores \citep{Croton_2006}. 
This is true not only for the secular evolution
of massive, virialized clusters; indeed, several strong radio sources are the beacon of massive protoclusters at redshift 2 and larger
\citep[see, e.g.,][]{Miley_2006,Hatch_2014,Chiaberge_2010, Overzier_2016, Tozzi_2022, Travascio_2025}.  
Recently, medium-deep JVLA observations of the core of the Spiderweb protocluster
in the S, X and Ka bands, showed in detail the interaction of powerful jets with the surrounding 
cold and hot medium \citep{Carilli_2022} and the ongoing magnetization process of the 
proto-ICM \citep{Anderson_2022}.
\\
\indent
Tracing the feedback in galaxy clusters in the radio band consists of
detecting the presence of radio AGN in the central galaxy, measuring
its total power and resolving the radio emission in the radio lobes
filling the cavities and driving the shock fronts in the ICM.  These studies have been very
successful for local targets, where deep radio and X-ray data are both
available, but they become increasingly difficult at higher redshifts.
Recently, a few studies focused on high-redshift clusters and on the
evolution of their properties with cosmic time.   
They show a lack of evolution in the fraction of clusters hosting cool cores \citep{AndradeSantos_2017, Ruppin_2021}, $\sim$40\% for $z\leq1.4$, and a similar stability over cosmic time in the fraction of clusters hosting cavities, $\sim$10\% for $z\leq1.1$ \citep{Hlavacek_2015,Olivares_2022}. 
It has thus
been realized that, at least out to $z$$\sim$1, the evolution of cool-cores mirrors the evolution of feedback.
Additionally, \citet{Santos_2010} showed that for low and medium redshift clusters, the radio
galaxies in the center of cool-cores have radio powers ranging
from $\sim$10$^{23}$ W/Hz to $10^{25}$ W/Hz.  
Only a few programs have been carried out to target medium and high redshift clusters 
with deep, high resolution radio data to perform a systematic
investigations of the radio-mode feedback up to $z$$\sim$1. Among these studies,
the JVLA follow-up  of CLASH \citep{Postman_2012} clusters \citep{Yu_2018}, and the investigation of the radio properties of BCGs in X-ray selected galaxy clusters \citep{Hogan_2015}, showed that 
$\geq$90\% of BCGs in relaxed clusters have 1.4~GHz radio powers above 10$^{23}$~W/Hz. In addition, high resolution X-ray data from {\sl Chandra} enable the measurement of the low-level 
nuclear activity in the BCG, despite the overwhelming ICM emission, 
and, therefore, the investigation of the complex relationship between radio and X-ray nuclear power
in BCGs \citep{Russell_2013, Yang_2018}. 
\\
\indent
Based on these considerations, we
conclude that in order to trace radio-mode AGN feedback we need to detect every radio galaxy at least down to
$10^{23}$ W/Hz (at 1.4 GHz) at high redshift.  In Figure
\ref{fluxlim.fig} we show the radio power in unresolved sources
corresponding to a detection limit of 1 $\mu$Jy (blue solid line).  Assuming that typically a detection requires a signal to
noise ratio $S/N \sim 5$, these values correspond to an rms noise of
0.2 $\mu$Jy per beam, 
which can be achieved with 30 hours of on-source observing time with the AA4 SKA-Mid telescope at Band 2.
This sensitivity is obtained at an angular resolution $\sim$0.75$''$ (Briggs 0, no tapering), below confusion limit, and as Figure~\ref{fluxlim.fig} demonstrates, it will allow a complete census of the radio
properties of the BCG above $10^{23}$ W/Hz at 1.4 GHz up to redshift $z\sim 2$ (the largest redshift where virialized galaxy clusters are currently found).
{A Deep Tier survey over about 10-30 deg$^2$, as the reference survey discussed by \citet{Prandoni-SKA}, could be completed in less than 2 months (see Table \ref{survey.tab}).
Considering that on the entire sky there are potentially $4\times
10^4$ massive clusters at $z>0.5$ which are within the reach of {\it NewAthena} or {\it AXIS}\footnote{\url{https://axis.umd.edu/}, expected launch date in 2032.} \citep{Reynolds_2024, Russell_2024, AXIS_2025}, the number of galaxies with radio activity in cool
cores at $z>0.5$ can be as high as $\sim 2 \times 10^4$, assuming
little or no evolution in the population of CC clusters, or few
thousands assuming a strong evolution by a factor of 10 \citep[so-far ruled out at least up to z$\sim$1, see][]{Ruppin_2021}.
Therefore, a strong synergy of the SKA with wide X-ray data will characterize the BCG in massive
clusters with a cool core and feedback activity and constrain the duty cycle of radio galaxies in
groups and clusters across the cosmic epochs.

\begin{SCfigure}[0.7][t]
\includegraphics[width=0.6\linewidth]{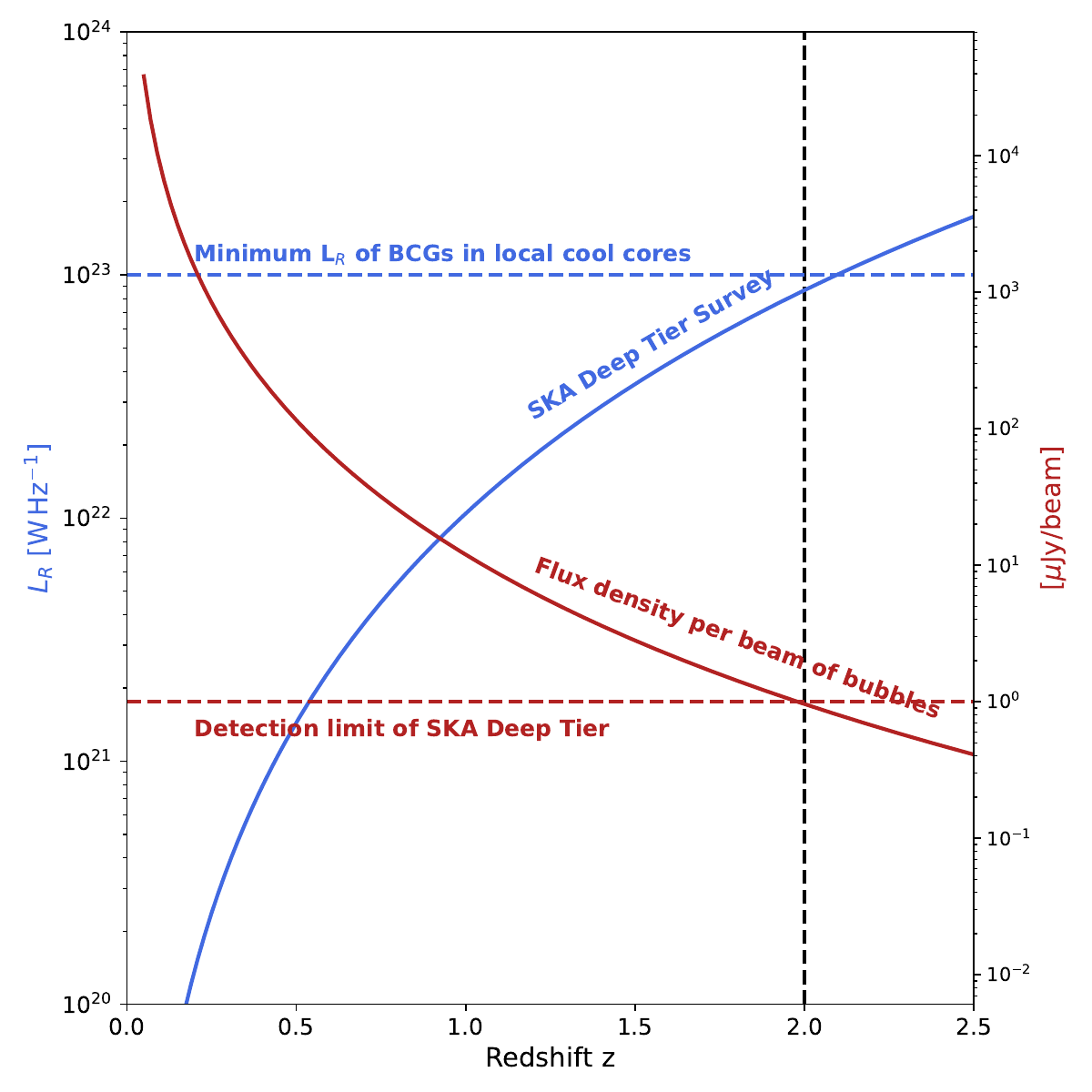}
\caption{\justifying{\it Left y-axis (blue):} 
The blue solid line shows the 1.4 GHz luminosity corresponding to a detection limit of 1~Jy (defined as $5 \times$ rms noise), which will be reached by SKA-Mid Deep Tier surveys (see Tab. \ref{survey.tab}). 
The blue horizontal dashed line indicates the minimum luminosity currently measured in radio galaxies hosted in local CC clusters.
{\it Right y-axis (red):} The red solid line shows the 
expected surface brightness
with a resolution of 0.75$''$ for a bubble with radio power of $\sim$10$^{24}$ W/Hz and physical size of $\sim$12 kpc, as in the reference cluster adopted in this work.  The red horizontal dashed line marks the detection limit of the Deep Tier surveys with SKA-Mid at the same angular resolution. The vertical dashed black line indicates the highest redshift where virialized, X-ray emitting clusters have been detected. 
  }\label{fluxlim.fig}
\end{SCfigure}

\subsection{Radio lobes, cavities and the energy budget of feedback}

A further major breakthrough in the investigation of radio-mode AGN
feedback at high redshift is the detection and characterization of the
radio lobes produced by the relativistic electrons, which are
responsible for carving the cavities in the ICM clearly observed
in X-ray local clusters \citep[see e.g.,][for a recent example]{Timmerman_2024}.  If we use as a reference the bubble size of RBS~797  (see Figure~\ref{RBS797feedback.fig}) we can 
  directly compute the typical angular size of the bubbles as a
function of the redshift.  
The
average size for a single RBS~797's bubble, assuming a spherical shape, is 12 kpc.
If we focus on targets at $z>1$ the angular size corresponds to about
1.5$''$, with a weak dependence on the
redshift.  We conservatively
assume that a beam FWHM of half the size of the bubble is sufficient
to resolve it.  Therefore, considering that the measured radio power at 1.4 GHz for a single cavity in RBS~797 is
 $10^{24}$ W/Hz  \citep[][]{Gitti_2006,Ubertosi_2024a}, we compute the radio flux density at 1.4
GHz from a single cavity as a function of redshift, assuming a
beam FWHM of 0.75$''$.
We show in Figure \ref{fluxlim.fig} that the signal from
a single cavity 
(red solid line) is above the
detection limit (defined as $5 \times$ rms noise) 
of AA4 SKA-Mid surveys with rms of 0.2 $\mu$Jy per beam (see Deep Tier survey in Table \ref{survey.tab})
up to $z \sim 2$. This indicates that the detection of AGN feedback signatures will rely on observations with the SKA telescopes, as the angular resolution of {\it NewAthena} ($\sim$9$''$) is not sufficient to detect X-ray cavities at high redshift. 

\subsection{Jet reorientation}

Ultimately, we consider the mechanism of jet reorientation and its relation to the distribution of feedback energy in cool cores. In some well-studied BCGs, dramatic misalignments between successive AGN outbursts have been reported. For example, in RBS~797, multiple AGN outbursts in different directions (up to $90^{\circ}$ misalignments) point to repeated changes in jet orientation \citep{Gitti_2006,Ubertosi_2021}. Explaining how jets reorient remains a major challenge, partially because of the lack of large samples. Existing investigations of RBS~797 relied on multi-frequency data from e.g., JVLA and LOFAR \citep[e.g.,][]{Gitti_2013a, Ubertosi_2024a}, to study the dynamics and timescales of jet activity (which requires data at $\geq$3 frequencies).  About $5\,\mu$Jy/beam sensitivity at $>$1 GHz with $\leq1''$ resolution was necessary to map the misaligned radio lobes. These requirements demonstrate why sample studies are difficult with current facilities. Yet systems like RBS~797 are not rare: 
$\sim$30\% of radio BCGs exhibit $\geq$45$^{\circ}$ jet reorientation \citep{Ubertosi_2024a}. Understanding this phenomenon is especially important for cool cores, where distributing feedback energy is essential to maintain thermal balance. 
 Observations with SKA-Mid at Band 2 will reach an rms of $\sim 1\,\mu$Jy/beam at $\sim $0.5$''$ resolution (Briggs $-1$, no tapering) in 3 hours on source. Therefore, a  Wide survey ($10^{3}$ square degrees, see Table \ref{survey.tab}) could be completed in about 4 months.
Using the same formalism of \cite{Gitti_2018}, we expect $\sim 2 \times 10^{3}$ X-ray luminous clusters over this area for $z\leq0.5$. Assuming a 40\% fraction of cool cores \citep[e.g.,][]{Ruppin_2021} and a 20\% fraction of cool cores with more than one pair of bubbles \citep[e.g.,][]{Olivares_2022}, we will be able to study the relative inclination of successive radio lobes in about $\sim$200 systems. Adopting a $30\%$ fraction of BCGs with $\geq45^{\circ}$ jet misalignments \citep[see][]{Ubertosi_2024b}, this implies $\gtrsim60$ clusters with strongly misaligned outbursts to follow-up with SKA-Mid pointed observations. With its sensitivities of 1-2 $\mu$Jy/beam in 1 hour at $\geq$1~GHz and resolutions down to 0.04$''$ \citep{SKAPerformance_2019}, SKA-Mid will achieve, in just $1.5$~hour per object (0.5 hour at three frequencies), the same level of detail that required $15$~hours of JVLA time for RBS~797. This will enable a systematic exploration of jet reorientation in radio galaxies at the center of cool cores. 
\vspace{-0.1in}

\section{Summary and conclusions}\label{conclusions.sec}
\vspace{-0.1in}

We explored the potential of future radio surveys with the full Design Baseline AA4 SKA-Mid telescopes to study the evolution of CC clusters. An outline of the surveys considered in this work is presented in Table \ref{survey.tab}.
\begin{table}[t]
	\centering
	\caption{Outline of the AA4 SKA-Mid surveys at Band 2 (central frequency $\nu=1.3$ GHz) considered in this work \citep[with performances similar to the reference surveys described in][]{Prandoni-SKA}. Column 1: Science drivers; column 2: survey tier; column 3: survey area in units of square degrees; column 4: on-source integration time in units of minutes; column 5: robustness of Briggs image weighting; column 6: tapering defined as arcsec in the image plane; column 7: continuum surface brightness sensitivity in units of $\mu$Jy/beam; column 8: major and minor axis (FWHM) of the synthesized beam-size in units of arcsec; column 8: survey duration in units of months, considering the expected SKA-Mid FOV ($\sim 1$ deg$^2$). All values are based on the current SKAO online sensitivity calculator for AA4 (\url{https://sensitivity-calculator.skao.int}).}
	\label{survey.tab}
	\begin{tabular}{lccccccccc}
		\hline
        \hline
		Science  & Tier & Area & time & R & Taper & rms & beam & duration\\
         & & (deg$^2$) & (min) & & $('')$ & ($\mu$Jy/b) & $('')$ & (months) \\
		\hline
        \hline
		Mini-halos & All-sky & $31 \times 10^3 $ & 10 & $0$ & 4.275 & 4.14 & 7.98$\times$6.95 & $\sim 7$ \\
        (BCG subtract.) & All-sky & $31 \times 10^3 $ & 10 & $0$ & no & 2.80 & 0.81$\times$0.69 & $\sim 7$\\
        \hline
        AGN feedback & Deep & 10 - 30 & 1800 & $0$ & no & 0.21 & 0.81$\times$0.69 & $<2$\\
        \hline
        Jet reorientation& Wide & $10^3$ & 180 & $-1$ & no & 1.11 & 0.48$\times$0.43 & $\sim 4$\\
        \hline
        \hline
\label{survey.tab}
\end{tabular}
\end{table}
Our predictions for radio mini-halos (Sect. \ref{followup.sec}), obtained by integrating the RLF of mini-halos derived from the XLF of CC clusters by adopting the observed $P_{1.4}$-$L_X$ correlation, suggest that SKA-Mid could identify up to approximately 3500 mini-halo candidates out to $z \sim 1$, although this relies on the optimistic assumption that all CC clusters host mini-halos.
If mini-halos originate from turbulent re-acceleration, their occurrence may decline in less massive or more distant clusters. Additionally, the synchrotron fraction of non-thermal emission decreases with redshift depending on the magnetic field strength, implying that SKA surveys could constrain the magnetic field properties via the evolving RLF of mini-halos.
For instance, based on our calculations, the SKA-Mid detection of more than 10 mini-halos at $z > 0.6$ would 
imply relatively strong magnetic fields ($B_{\rm MH} \geq 10$-$30 \, \mu$G), 
whereas a non-detection at the same redshift would point to much weaker fields ($B_{\rm MH} < 1 \, \mu$G). Possible challenges in the detection of radio mini-halos include contamination from bright point sources, especially central BCGs, and limitations in dynamic range. However, SKA’s planned angular resolutions, along with the possibility to implement visibility-based subtraction techniques, 
should mitigate these issues. 
\\
We also highlighted the synergy between radio surveys and future X-ray missions (Sect. \ref{athena.sec}).  Simulations indicate that turbulence with velocity broadening $\sigma_{\rm turb} \sim$ 50-200 km/s on $\lesssim$100 kpc scales can re-accelerate relativistic electrons and produce mini-halos. These turbulence levels are within reach of {\it NewAthena}'s X-IFU, which will allow a follow-up of $>50$\% of SKA-Mid samples up to $z \sim 1$ by detecting anisotropic turbulence down to tens of km/s on arcsecond scales.

Our feasibility study further showed that the
radio-mode AGN feedback will be unveiled practically at any level in
clusters up to $z \sim 2$,  close to the maximum redshift where virialized clusters have been detected in the X-ray band so far (Sect. \ref{feedback.sec}). 
When the SKAO will be fully operational, the knowledge of distant galaxy
clusters will be largely but unpredictably changed.  A large number of
optically and IR selected clusters will be available thanks to the
forthcoming surveys of the Euclid\footnote{\url{http://sci.esa.int/euclid/}.}
satellite and of the Vera C. Rubin Observatory\footnote{\url{https://rubinobservatory.org/}.}.
Thanks to the ICM X-ray characterization of the cluster cores
achievable with the {\it NewAthena} mission, and potentially the {\it AXIS} telescope, these can be directly combined
with the radio data to fully characterize jet-driven bubbles and feedback.

\bibliographystyle{abbrvnat-maxbibnames4}
\bibliography{bibliography-SKAgitti} 

\end{document}